\documentclass[aps,prl,twocolumn,showpacs,superscriptaddress]{revtex4-2}
\usepackage{graphicx}
\usepackage{amsmath}
\usepackage{multirow}
\usepackage{tabularx}
\usepackage{color}
\usepackage{xcolor}
\usepackage{ulem}
\usepackage[colorlinks,linkcolor=blue,urlcolor=blue,anchorcolor=blue,citecolor=blue]{hyperref}
\begin{document}
\preprint{APS/123-QED}
\title{Progress of ambient-pressure superconductivity in bilayer nickelate thin films}

\author{Wenyuan Qiu}
\author{Dao-Xin Yao}
\email{yaodaox@mail.sysu.edu.cn}
\affiliation{Guangdong Provincial Key Laboratory of Magnetoelectric Physics and Devices, State Key Laboratory of Optoelectronic Materials and Technologies, School of Physics, Sun Yat-Sen University, Guangzhou, 510275, China}

\begin{abstract}
This review summarizes recent progress of ambient-pressure superconductivity in bilayer nickelate La$_3$Ni$_2$O$_7$ thin films, a major advancement following the discovery of high-pressure superconductivity in bulk La$_3$Ni$_2$O$_7$. 
First, we explain how epitaxial strain engineering enables ambient-pressure superconductivity in La$_3$Ni$_2$O$_7$ thin films, with compressive strain from substrates like SrLaAlO$_4$ stabilizing superconductivity. 
Next, we review experimental characterizations of related systems, with particular emphasis on ARPES measurements that have shown conflicting Fermi surface topologies. 
We then discuss progress in increasing the superconducting transition temperature $T_c$. 
Finally, we summarize theoretical studies of the electronic structure and pairing symmetry of La$_3$Ni$_2$O$_7$ thin films.
Together, these advances establish bilayer nickelate thin films as a highly tunable and promising platform for exploring high-$T_c$ superconductivity.
\end{abstract}

\maketitle

\section{\label{sec:1}Introduction}
Since the discovery of superconductivity (SC) in copper oxides with a transition temperature ($T_c$) of up to 164 K~\cite{Bednorzpossible1986,Gaosuperconductivity1994}, many efforts have been devoted to finding more high-$T_c$ superconductors.
One of the promising candidates predicted to be a high-$T_c$ superconductor is nickel oxides, as nickel is located next to copper in the periodic table and is therefore believed to possess similar properties~\cite{Anisimovelectronic1999}.
In 2019, the prediction was finally confirmed with the discovery of SC in an infinite-layer nickelate~\cite{li2019superconductivity}.
In 2023, Sun \textit{et al.}~\cite{Sunsignatures2023} found SC with $T_c$ near 80 K in bilayer nickelate La$_3$Ni$_2$O$_7$ under high pressure, marking nickelates as the second class of high-$T_c$ superconductor to enter the liquid-nitrogen temperature range, following copper oxides.
In 2024, Zhu \textit{et al.}~\cite{Zhusuperconductivity2024,Lisignature2024,Zhangsuperconductivity2025} observed SC with $T_c$=20-30 K under high pressure in trilayer nickelate La$_4$Ni$_3$O$_{10}$, further expanding the family of nickelate superconductors.
However, the requirement of high pressure has significantly limited the use of specific key experimental techniques for directly investigating the superconducting phase.
Consequently, considerable research has focused on achieving ambient-pressure SC in nickelates.

\begin{figure}[b]
    \centering
    \includegraphics[width=1\linewidth]{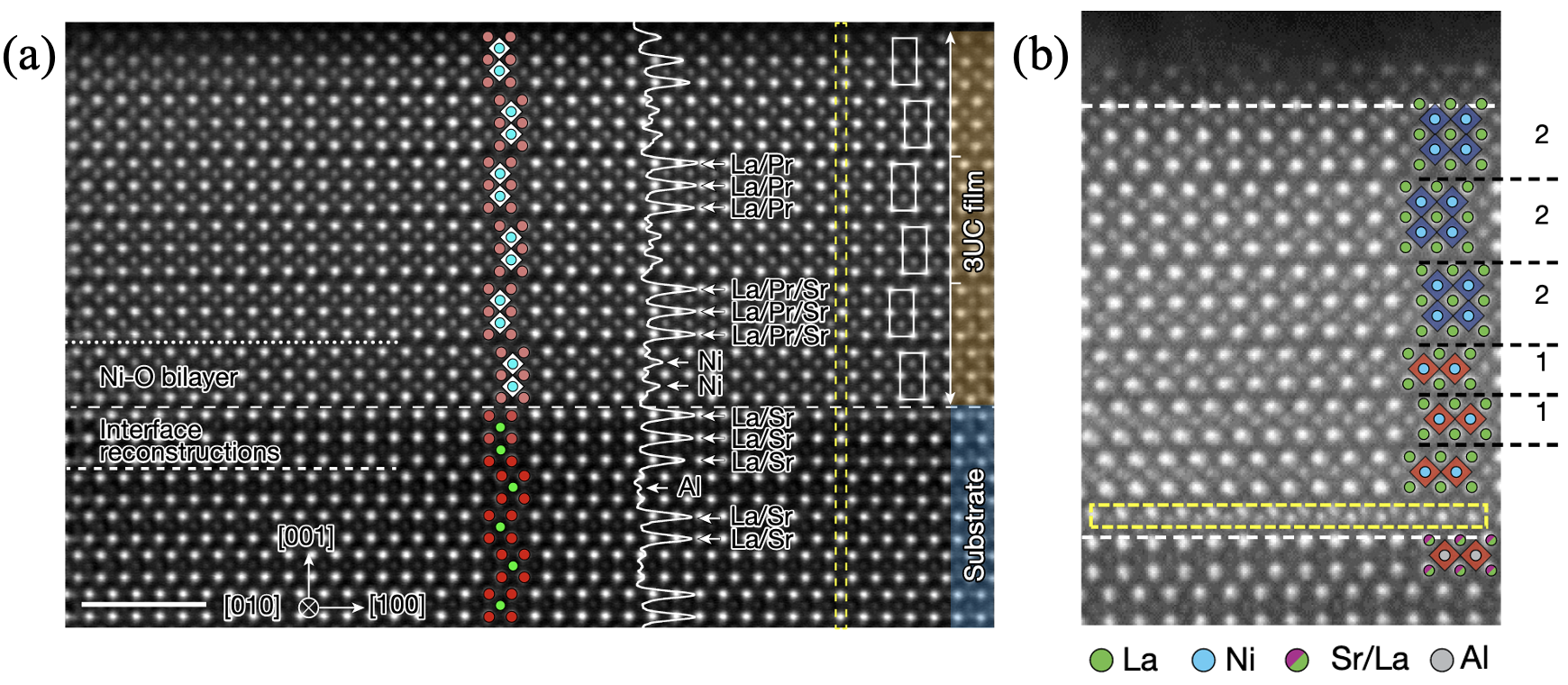}
    \caption{Microscopic structures of (a)  La$_{2.85}$Pr$_{0.15}$Ni$_2$O$_7$ sample on SrLaAlO$_4$~\cite{Zhouambient-pressure2025} and (b) of  La$_3$Ni$_2$O$_7$ on SrLaAlO$_4$~\cite{Kosignatures2025}.}
    \label{fig:1-1}
\end{figure}

Recently, using pulsed laser deposition (PLD),
Ko \textit{et al.}~\cite{Kosignatures2025} reported ambient-pressure SC with a $T_c$ exceeding 40 K in La$_3$Ni$_2$O$_7$ thin films grown on SrLaAlO$_4$ (SLAO) substrates. 
Almost simultaneously, Zhou \textit{et al.}~\cite{Zhouambient-pressure2025} independently observed ambient-pressure SC with a $T_c$ over 40 K in La$_{2.85}$Pr$_{0.15}$Ni$_2$O$_7$ thin films grown on the same substrates {\color{black} by gigantic-oxidative  atomic-layer-by-layer epitaxy (GAE)}~\cite{KanaiLow1989,zhougigantic2025}.
{\color{black} And many experiments have been conducted on the samples synthesized by GAE~\cite{Liangle2025,Nieambient2025,Zhousuperconductivity2025}.}
The microscopic structures of these thin films are shown in Fig.~\ref{fig:1-1}.
Later, several groups reported ambient-pressure SC in related systems, {\color{black} including} La$_2$PrNi$_2$O$_7$~\cite{Liusuperconductivity2025,Wangelectronic2025,Fansuperconducting2025}, and La$_{3-x}$Sr$_x$Ni$_2$O$_7$ thin films~\cite{Haosuperconductivity2025, Sunobservation2025,Zhongdoping2026}.
{\color{black} And more and more La$_3$Ni$_2$O$_7$ thin-film samples are synthesized and various experimental explorations have been conducted on them until right now}~\cite{Bhattresolving2025,Osadastrain2025,Shennodeless2025,Jitime2026}.
{\color{black} And among these findings, the effect of element doping on La$_3$Ni$_2$O$_7$ thin films has been investigated systematically.
Partial substitution of La with Pr in La$_3$Ni$_2$O$_7$ has been reported to improve sample quality, enhancing phase purity and SC~\cite{Zhouambient-pressure2025,Liusuperconductivity2025}.
The effect of carrier doping is studied through Sr$^{2+}$ doping in La$_{3-x}$Sr$_x$Ni$_2$O$_7$, which serves as an equivalent approach for hole doping, giving a phase diagram where the highest $T_c$ value of $~42$ K is achieved at $x=0.21$~\cite{Haosuperconductivity2025}.
}
These findings represent a significant advancement in the field of nickelate superconductors.

In this review, we begin by examining the role of epitaxial strain in enabling ambient-pressure SC in La$_3$Ni$_2$O$_7$ thin films. 
We then examine key experimental studies, especially ARPES measurements that reveal distinct Fermi surface (FS) topologies, and summarize methods for raising the superconducting transition temperature $T_c$. 
Finally, we review recent theoretical studies of the electronic structures and possible pairing symmetry in La$_3$Ni$_2$O$_7$ thin films.

\begin{figure*}[t]
    \centering
    \includegraphics[width=1\linewidth]{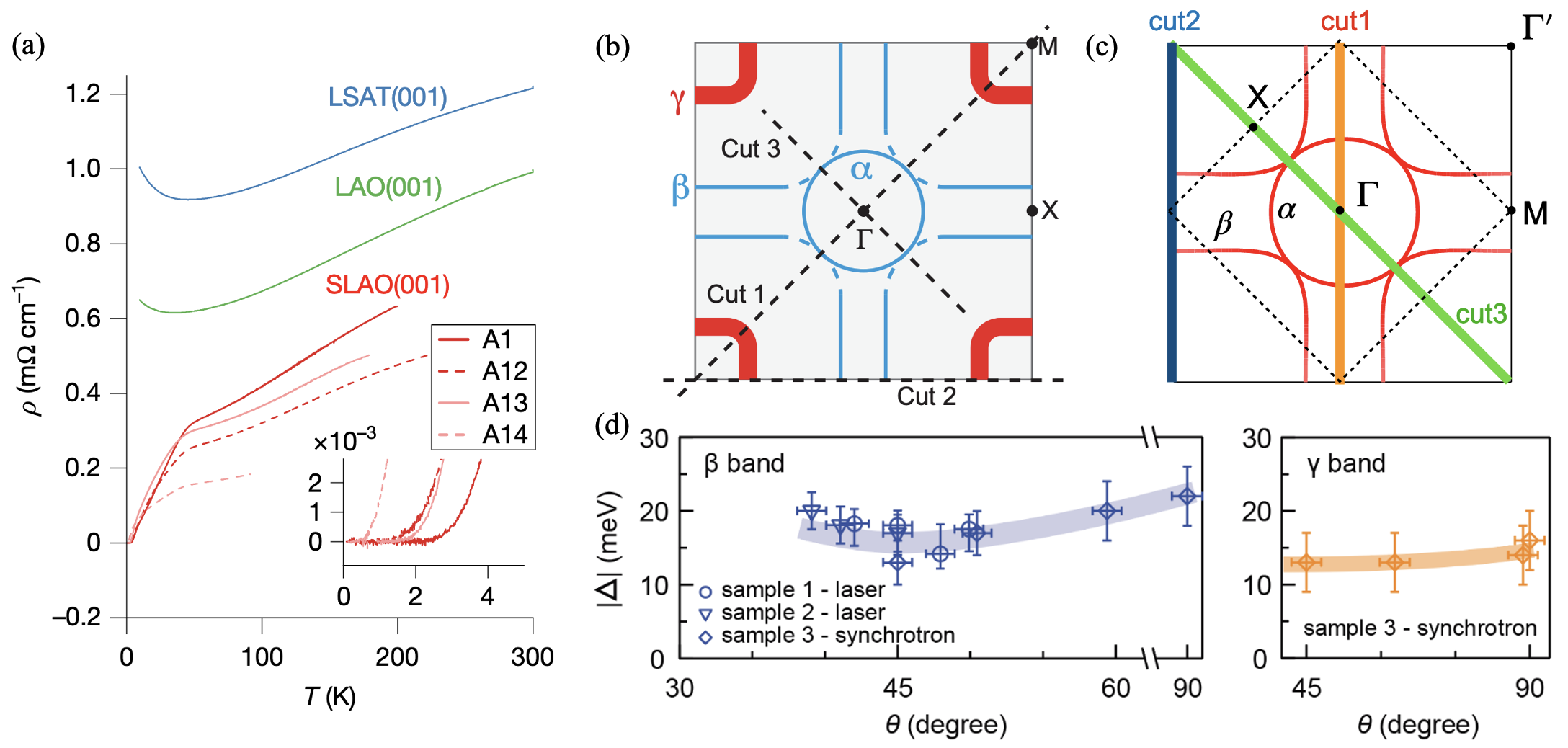}
    \caption{(a) Temperature-dependent resistivity $\rho$(T) of  La$_3$Ni$_2$O$_7$ thin films grown on different substrates~\cite{Kosignatures2025}. (b)FS of La$_{2.85}$Pr$_{0.15}$Ni$_2$O$_7$ thin films, consisting of $\alpha$, $\beta$ and $\gamma$ pockets~\cite{Liangle2025}. (c) FS of La$_{2}$PrNi$_2$O$_7$ thin films, without $\gamma$ pocket appearing on FS~\cite{Wangelectronic2025}. (d) Measured superconducting gap magnitudes on the $\beta$ and $\gamma$ bands as a function of the angle $\theta$~\cite{Shennodeless2025}.}
    \label{fig:1}
\end{figure*}

\section{\label{sec:2}Effects of epitaxial strain}

To achieve ambient-pressure SC, one promising approach is to apply epitaxial strain to La$_3$Ni$_2$O$_7$ thin films.
Before discussing the effects of epitaxial strain, it is necessary to analyze the pressure phase diagram of bulk La$_3$Ni$_2$O$_7$~\cite{Liidentification2025}, which can be divided into a low-pressure region (LP) and a high-pressure region (HP).
The characteristic structure of La$_3$Ni$_2$O$_7$ is its bilayer NiO$_2$ planes, where each layer contains a NiO$_6$ octahedron ~\cite{Wangnormal2024, Wangrecent2025}. 
The two octahedra share an apical oxygen, forming a Ni-O-Ni bond.
In the LP phase, the space group is $Aman$, characterized by a 168$^\circ$ bond angle.
As the pressure approaches approximately 10 GPa, it transitions into the HP phase, characterized by two octahedra aligned with each other, leading to a bond angle tilting to 180$^\circ$ and the space group changing to $Fmmm$ above 14 GPa and $I4/mmm$ above 46.8 GPa.

Since SC only appears in the HP phase, the main challenge is to stabilize this HP phase in thin films at ambient pressure.
Cui \textit{et al.}~\cite{Cuistrain2024} revealed that epitaxial misfit strain is the dominant factor controlling the phase formation in RP nickelates La$_{n+1}$Ni$_n$O$_{3n+1}$.
Their report indicates that tensile strain stabilizes the perovskite LaNiO$_3$ (n=$\infty$) phase, while compressive strain favors the formation of the La$_3$Ni$_2$O$_7$ (n=2) phase.
Therefore, substrates play an important role in the induction of SC in bilayer nickelate thin films.
As shown in Fig.~\ref{fig:1} (a), the temperature-dependent resistivity $\rho(T)$ of La$_3$Ni$_2$O$_7$ thin films grown on SLAO substrates begins to decrease at approximately 42 K and reaches zero resistance near 2 K.
In contrast, $\rho(T)$ for films grown on LaAlO$_3$ (LAO) and (LaAlO$_3$)$_{0.3}$(Sr$_2$TaAlO$_6$)$_{0.7}$ (LSAT) substrates shows an upturn below 40 K without reaching a zero-resistance state~\cite{Kosignatures2025}.
The main difference among these substrates comes from their in-plane lattice constants, with SLAO applying compressive strain, LAO exerting only a mild compressive strain, and LSAT introducing a slight tensile strain.
These results demonstrate that compressive strain is crucial for inducing ambient-pressure SC in La$_3$Ni$_2$O$_7$ thin films.
In addition, the results from Chen's group showed that the transition to zero resistance exhibits signatures of a Berezinskii–Kosterlitz–Thouless (BKT) transition~\cite{Zhouambient-pressure2025}.

\section{\label{sec:3}characterization results}
\begin{figure}[t]
    \centering
    \includegraphics[width=1\linewidth]{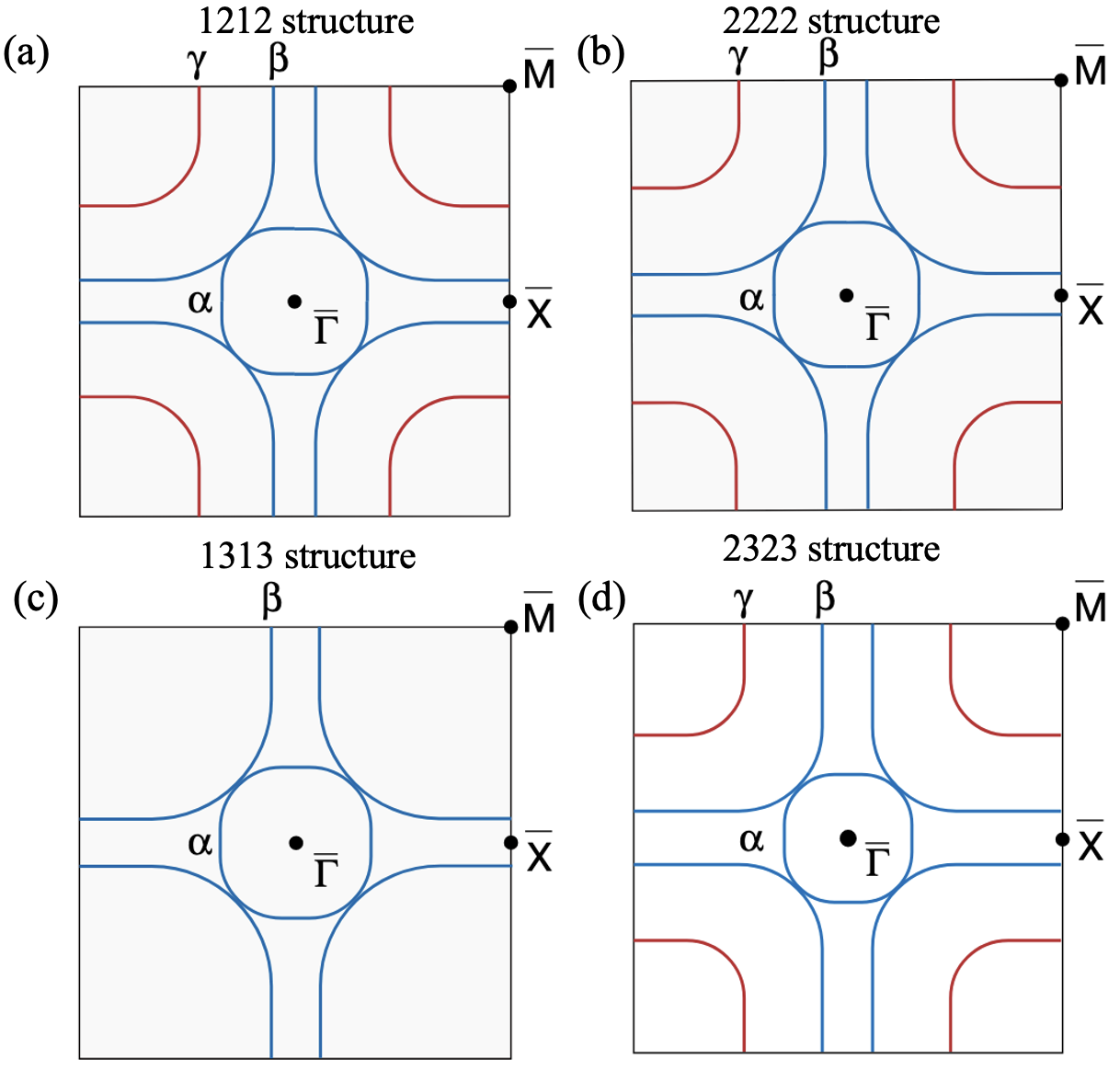}
    \caption{FS~\cite{Nieambient2025} of thin films grown on SLAO substrates measured by ARPES for (a) 1212 structure, (b) 2222 structure, (c) 1313 structure and (d) 2323 structure.}
    \label{fig:2}
\end{figure}

The absence of high-pressure requirements in these thin films enables direct experimental investigation of the superconducting phase. 
Consequently, various measurements have been performed on bilayer nickelate thin films to investigate the underlying mechanisms of SC.
Scanning transmission electron microscopy (STEM) measurements show that the apical Ni–O–Ni bond angle approaches 180$^\circ$~\cite{Kosignatures2025}, which is very similar to the HP phase of bulk La$_3$Ni$_2$O$_7$.
{\color{black} It is also notable that ozone annealing plays an important role in ambient SC of La$_3$Ni$_2$O$_7$ thin films.
After O$_3$ annealing, the as-grown La$_3$Ni$_2$O$_7$ thin films transition from an insulating state into a superconducting state~\cite{Kosignatures2025}.
And the X-ray absorption spectroscopy (XAS) measurements further indicate that the O$_3$-annealed thin films contain a combination of 50\% Ni$^{2+}$ and 50\% Ni$^{3+}$, whereas the as-grown thin films contain 55\% Ni$^{2+}$ and 45\% Ni$^{3+}$~\cite{Kosignatures2025}.
Thus, the Ni ion in the O$_3$-annealed thin films exhibits a mixed valence state of 2.5.
These facts show that ozone annealing can increase oxygen content, and maintain a Ni$^{2.5+}$ valence state in La$_3$Ni$_2$O$_7$ thin films.}
Angle-resolved photoemission spectroscopy (ARPES) studies~\cite{Liangle2025,Shennodeless2025,Wangelectronic2025} on bilayer nickelate thin films have been conducted to confirm the topology of FS.
As shown in Fig.~\ref{fig:1} (b), ARPES measurements on La$_{2.85}$Pr$_{0.15}$Ni$_2$O$_7$ thin films~\cite{Liangle2025} reveal a FS, where the $\alpha$, $\beta$ and $\gamma$ pockets are identified.
Furthermore, direct measurement of the superconducting gap on the FS~\cite{Shennodeless2025} reveals a significant gap opening on the $\beta$ FS sheet with no signs of nodes along the Brillouin-zone diagonal, as shown in Fig.~\ref{fig:1} (d).
{\color{black} Notably, this gap survives to a temperature above $T_c$ and shows a particle–hole symmetric evolution, consistent with the presence of a pseudogap.}
However, ARPES measurements on La$_{2}$PrNi$_2$O$_7$ thin films~\cite{Wangelectronic2025} give a FS without the $\gamma$ pocket, which is about 70 meV below the Fermi level, as shown in Fig.~\ref{fig:1} (c).
{\color{black} ARPES measurements on Sr-doped La$_3$Ni$_2$O$_7$ thin films also reports a FS without the $\gamma$ pocket~\cite{Sunobservation2025}.}
The contrast results may arise from differences in the growth conditions {\color{black} and measurement environment}  of the thin films.
{\color{black} It has been reported that Sr diffusion from substrates was observed, which may also result in the appearance of the $\gamma$ pocket~\cite{Liangle2025}.
Whether $\gamma$ pocket appears in the intrinsic La$_3$Ni$_2$O$_7$ thin films under 2\% compressive strain is still under debate.}
Scanning tunneling microscopy (STM) measurements~\cite{Fansuperconducting2025} reveal a two-gap structure on the FS, with fitting analyses indicating a preferred anisotropic $s$-wave pairing. 
Taken together, these observations support a predominant $s\pm$-wave pairing symmetry in the system.
But whether the pairing symmetry is $s\pm$ wave is still under debate.
Most recently, Nie \textit{et al.}~\cite{Nieambient2025} reported ambient-pressure SC with onset $T_c$ up to 50 K in both hybrid monolayer-bilayer (1212) and pure bilayer (2222) films, and onset $T_c$ of 46 K in bilayer–trilayer (2323) thin films, while the hybrid monolayer-trilayer (1313) structure remained non-superconducting. 
{\color{black} The FS of these films measured by ARPES in Fig.~\ref{fig:2} showed a clear difference: a hole-like $\gamma$ band crosses the Fermi level in the superconducting films, but in the non-superconducting 1313 film, it is below the Fermi level.
A recent theoretical work attributes the suppression of SC in 1313 La$_3$Ni$_2$O$_7$ to reduced pairing strength in the trilayer subsystem and weakened phase coherence between trilayer subsystems arising from S–N–S Josephson coupling~\cite{Chenpairing2026}.}

\section{\label{sec:4}Methods to increase $T_c$}
Following the discovery of SC in bulk and thin-film La$_3$Ni$_2$O$_7$, methods to raise $T_c$ have emerged as a major research focus.
In thin film systems, methods to increase $T_c$ mainly {\color{black} involve enhancing thin-film growth techniques and applying pressure}.
Zhou \textit{et al.}~\cite{Zhousuperconductivity2025} showed that their improved GAE technique, achieved by pushing the growth regime into an extreme non-equilibrium state,  can stabilize an ambient-pressure superconducting phase with an onset temperature of up to 63 K.
{\color{black} Osada \textit{et al.}~\cite{Osadastrain2025} showed that as the $c/a$ ratio increases, the onset $T_c$ under hydrostatic pressure rises from approximately 10 K under tensile strain to nearly 60 K under compressive strain.
}
{\color{black} Applying hydrostatic pressure on compressively strained La$_3$Ni$_2$O$_7$ thin films, which raises the onset $T_c$ to over 60 K, has also been reported recently~\cite{Lienhanced2025}.}
{\color{black} And most recently, Zhao \textit{et al.}~\cite{Zhaopressure2026} showed that hydrostatic pressure universally enhances superconductivity in (La,Pr)$_3$Ni$_2$O$_7$ thin films, raising the onset $T_c$ to 68.5~K at 2.0~GPa. }
{\color{black} They attribute the observed resistance dip above $T_c$ to oxygen-vacancy-induced electron localization. 
The dip is suppressed by pressure, which directly correlates with the increase in $T_c$, establishing oxygen vacancies as a key tuning parameter for SC in bilayer nickelates.}
{\color{black} Overall, $T_c$ in thin films is probably enhanced by pressure via increasing interlayer coupling, orbital hybridization, and spin fluctuations.}
Also, some theoretical works~\cite{Shaopossible2026,Fanminimal2025} have explored methods to enhance $T_c$ by applying an electric field to thin films, which require experimental verification.
While in bulk systems, using element substitution to increase $T_c$~\cite{Paneffect2024} has advanced significantly.
In bulk systems, Li \textit{et al.}~\cite{Libulk2025} reported bulk SC in La$_2$SmNi$_2$O$_7$ under high pressure, which exhibits a $T_c$ up to 96 K, zero resistance temperature up to 73 K, and clear Meissner screening, confirming robust bulk high-$T_c$ behavior.
Qiu \textit{et al.}~\cite{Qiuinterlayer2025} demonstrated that Nd substitution in bilayer La$_3$Ni$_2$O$_7$ compresses the lattice and enhances interlayer magnetic coupling, resulting in SC with an onset $T_c$ of up to approximately 98 K.
Chen \textit{et al.}~\cite{Chenelectronic2025} systematically calculated the electronic structures of Nd-doping bulk La$_3$Ni$_2$O$_7$, revealing that increasing Nd doping leads to a larger interlayer $d_{z^2}$ orbital hopping, which would result in a larger interlayer superexchange coupling and a higher $T_c$.
{\color{black} In conclusion, rare-earth element substitutions can induce chemical pressure in bulk La$_3$Ni$_2$O$_7$ to raise $T_c$.
It has been reported that applying pressure can increase $T_c$ in thin films~\cite{Lienhanced2025}, suggesting that $T_c$ might be enhanced by chemical pressure.
The effectiveness of this approach needs further experimental verification.}

\begin{figure}[t]
    \centering
    \includegraphics[width=1\linewidth]{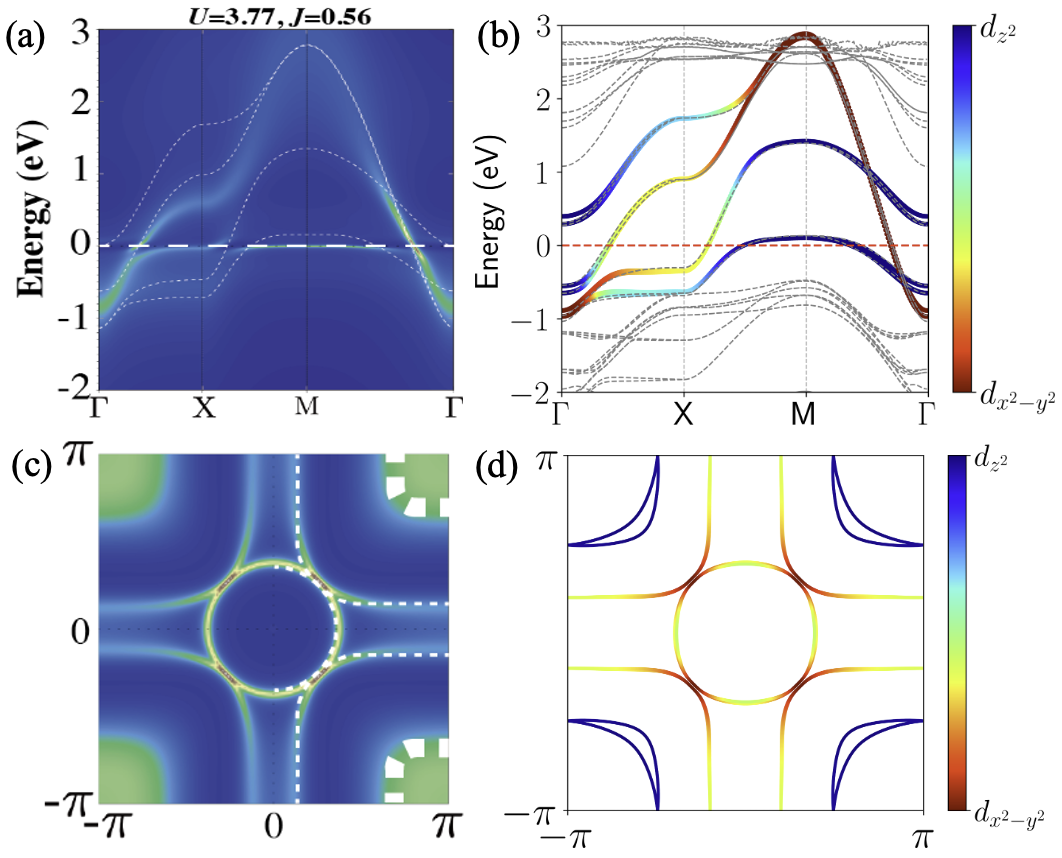}
    \caption{(a) The energy bands and (c) FS of La$_3$Ni$_2$O$_7$ thin films, combining DFT and DMFT~\cite{Yuecorrelated2025}. (b) The energy bands and (d) FS of the One-UC double-stack tight-binding model for La$_3$Ni$_2$O$_7$ thin films~\cite{Huelectronic2025}.}
    \label{fig:3}
\end{figure}

\begin{figure}[t]
    \centering
    \includegraphics[width=1\linewidth]{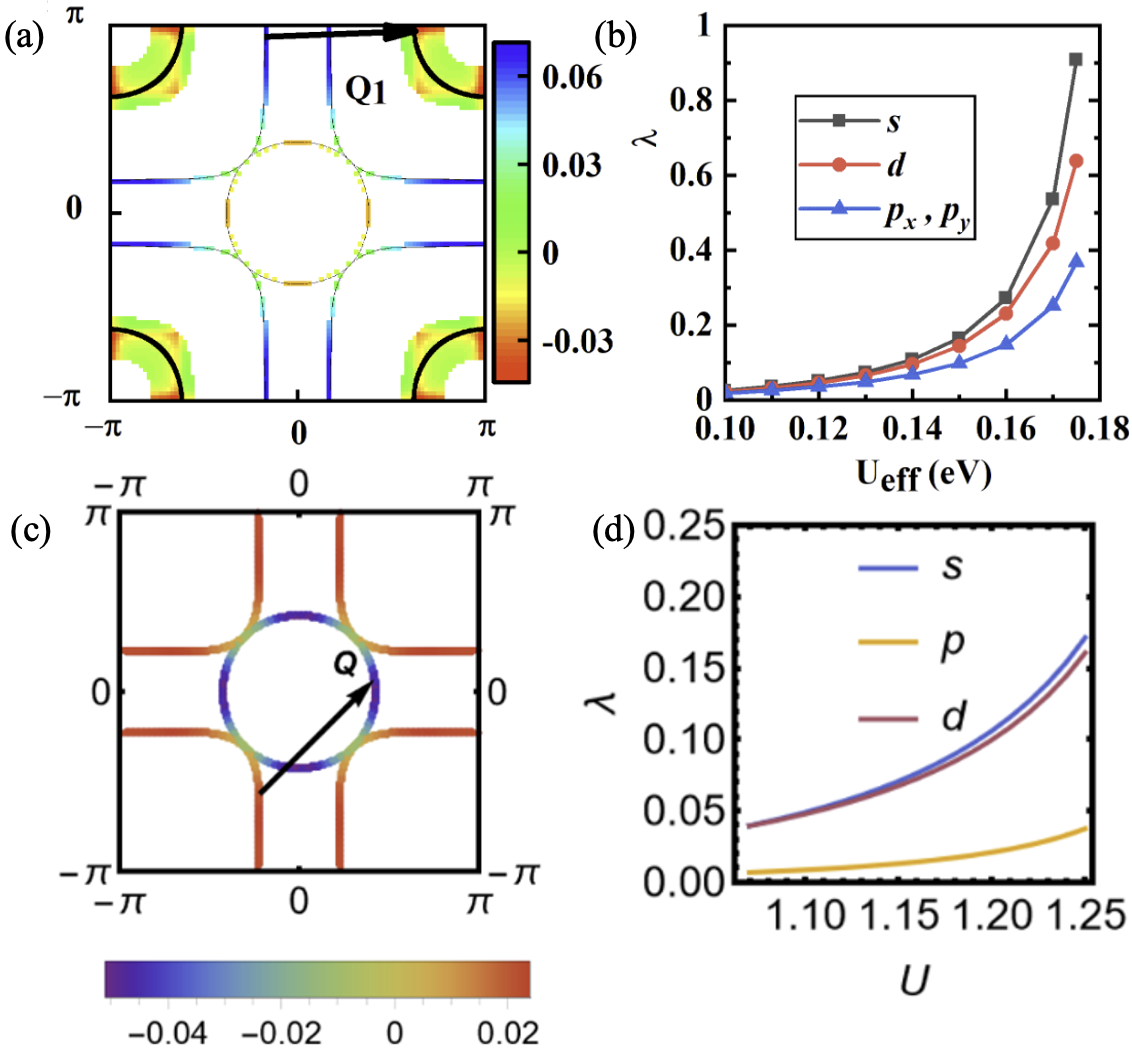}
    \caption{(a) The leading cRPA calculated gap functions on the FS and (b) the dependence of $\lambda$ on $U_{eff}$~\cite{Yuecorrelated2025}. (c) The leading RPA calculated gap functions on the FS and (d) the dependence of $\lambda$ on $U$ without $\gamma$ pocket~\cite{Shaopairing2025}.}
    \label{fig:4}
\end{figure}

\begin{figure*}[t]
    \centering
    \includegraphics[width=1\linewidth]{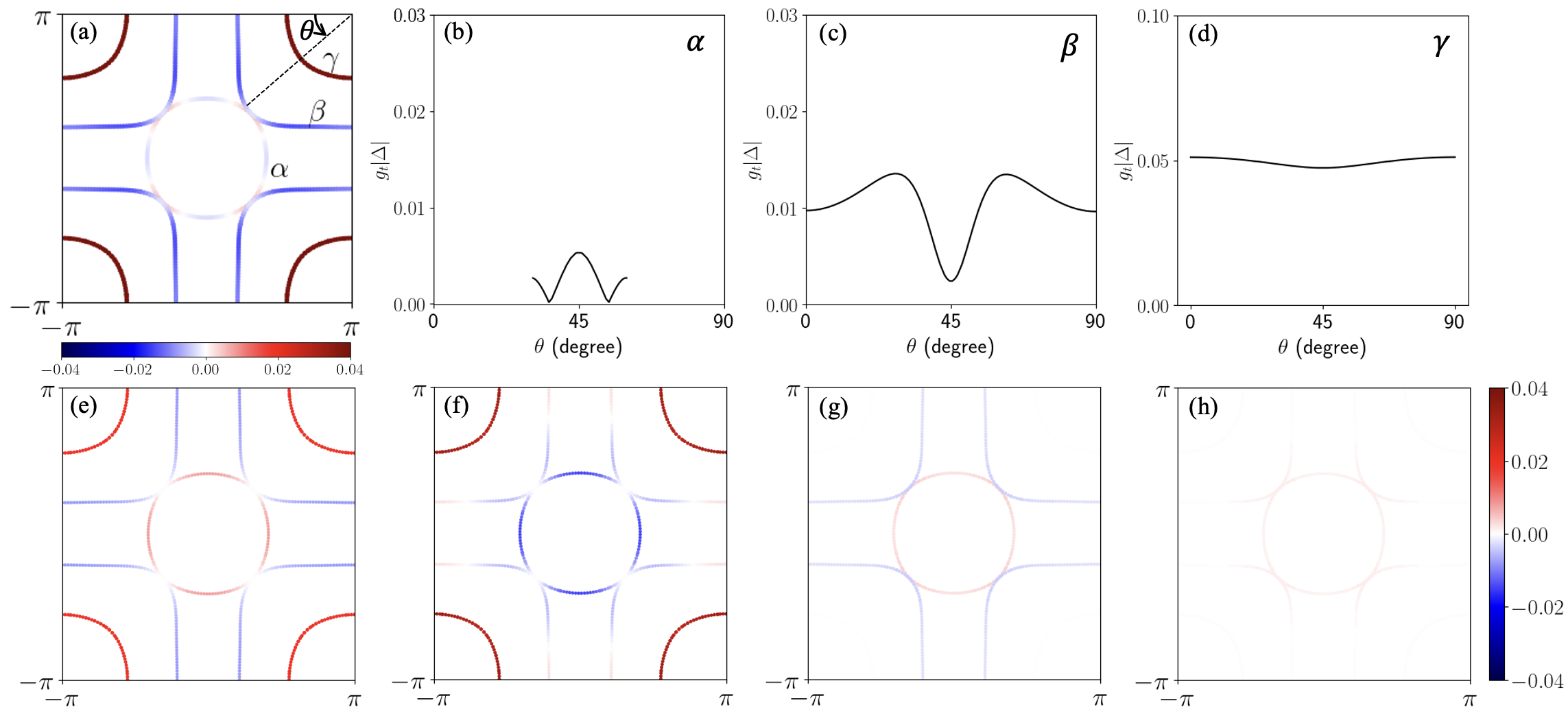}
    \caption{(a) Superconducting gap projected on the FS and (b)-(d) superconducting gap as a function of angle for $\alpha$, $\beta$ and $\gamma$ pockets, respectively.
    (e)-(h) Orbital-resolved superconducting gap on the FS for interlayer $d_{z^2}$, inplane $d_{z^2}$, interlayer $d_{x^2-y^2}$ and inplane $d_{x^2-y^2}$ pairing, respectively~\cite{Qiupairing2025}.}
    \label{fig:5}
\end{figure*}

\section{\label{sec:5}Theory progress of bilayer nickelate thin films}
Since the discovery of SC in bilayer nickelate thin films, significant focus has been directed toward their underlying electronic structures. 
Several research groups subsequently performed systematic calculations of the electronic band structures~\cite{Yuecorrelated2025,Huelectronic2025, Leopposite2025, Shieffect2025, Liusuperconductivity2025, Bhattresolving2025}, employing first-principles density functional theory (DFT) and model analyses to characterize the low-energy states, FS topology, and orbital composition~\cite{Luobilayer2023}.
Using the constrained random phase approximation (cRPA), Yue \textit{et al.}~\cite{Yuecorrelated2025} demonstrated intra-orbital Coulomb interaction $U\approx3.77$ eV and Hund's coupling $J_H\approx0.56$ eV.
Combining DFT and dynamical mean-field theory (DMFT) with the $U$ and $J_H$ obtained from cRPA and a particle filling of $n=1.3$, they reproduced FS comparable to ARPES results, as shown in Fig.~\ref{fig:3} (a) and (c).
Using DFT, Hu \textit{et al.} ~\cite{Huelectronic2025} systematically investigated the electronic structures and slab models across various thicknesses.
They constructed the One-UC (unit cell) double-stack tight-binding model and the Half-UC slab model, which is used for comparison, and proposed a double-stack high-energy $d$–$p$ model for the first time, laying the groundwork for future research.
The energy bands and FS of the One-UC double-stack tight-binding model are shown in Fig.~\ref{fig:3} (b) and (d).
Since the One-UC slab consists of two bilayers, there is some small hopping between the two bilayers, resulting in a small split of the $\gamma$ pocket.
It is notable that the interlayer $d_{z^2}$ orbital hopping parameter for the One-UC slab is -0.550, while for the Half-UC slab it is -0.503. 
This suggests that the One-UC slab may have a larger interlayer $d_{z^2}$ orbital superexchange coupling compared to the Half-UC case, according to $J\sim 4t^2/U$.
Based on the double-stack model, the random phase approximation (RPA) spin susceptibility exhibits the strongest response reflecting nesting intra $\gamma$ pocket.
Building on a similar double-stack model, Li~\textit{et al.}~\cite{Litheoretical2025} performed a DFT+$U$ study of La$_3$Ni$_2$O$_7$/SrLaAlO$_4$ thin films, explicitly incorporating both substrate-induced compressive strain and interfacial Sr interdiffusion.
Recently, using DQMC and DMFT, Zhong \textit{et al.}~\cite{Zhongsuperexchanges2025} reported that, La$_3$Ni$_2$O$_7$ thin films exhibit a significantly enhanced charge-transfer capability together with interlayer and intralayer antiferromagnetic correlations of comparable magnitude.

Pairing symmetry is an important question in high-$T_c$ superconductors. 
Weak-coupling methods, such as RPA~\cite{Liuthes2023, Zhangstructural2024} and functional renormalization group (FRG)~\cite{Yangpossible2023, Jiangtheory2025}, are typically used to find out the pairing symmetry.
Yue \textit{et al.}~\cite{Yuecorrelated2025} used a modified RPA approach based on their model, resulting in an $s\pm$ wave pairing symmetry, as shown in Fig.~\ref{fig:4} (a) and (b).
However, the RPA approach is very sensitive to the details of FS.
Shao \textit{et al.}~\cite{Shaopairing2025} {\color{black} constructed} a tight-binding model without $\gamma$ pocket, giving rise to an $s\pm$ wave pairing symmetry, driven by nesting between the $\alpha$ pocket and $\beta$ pocket, as shown in Fig.~\ref{fig:4} (c) and (d).
{\color{black} A work using the variational Monte Carlo method also reports a robust $s\pm$ symmetry against change in FS~\cite{Watanabehierarchical2026}.}
Combining first-principles and FRG calculations, Le \textit{et al.}~\cite{Leopposite2025} found that scattering between FS sheets with opposite parity symmetry enhances interlayer $s\pm$ wave SC, while nesting between FS sheets with the same parity symmetry would break the pairing.
Using FRG, Cao~\textit{et al.}~\cite{Caostrain2026} found that the pairing symmetry for both La$_3$Ni$_2$O$_7$ and La$_{2.85}$Pr$_{0.15}$Ni$_2$O$_7$ thin films is $s\pm$ wave and $T_c$ could be enhanced under in-plane compression.
Recently, Zhang \textit{et al.}~\cite{Zhangcompressive2025} used the RPA approach on a one-UC compressive La$_3$Ni$_2$O$_7$ thin film, revealing a leading $d_{x^2-y^2}$ pairing state at moderate hole doping, and a $d_{xy}$ pairing symmetry with higher doping level.
{\color{black} However, there are also some works supporting $d_{xy}$ wave or $d+is$ symmetry~\cite{Xiasensitive2025,Zhensuperconductivity2024}.
And a recent measurement of differential  conductance $(dI/dV)$ spectra on pressurized La$_3$Ni$_2$O$_{7-\delta}$ single crystal suggests a $d$-wave-like symmetry~\cite{Caodirect2025}. 
}
Another theoretical method for identifying the pairing symmetry is the renormalized mean-field theory (RMFT)~\cite{Luohigh2024,Zhangrenormalised1988}, which begins with the $t-J$ model.
Qiu \textit{et al.}~\cite{Qiupairing2025} used the $t-J$ Hamiltonian, including superexchange couplings $J^z_{\bot}$, $J^x_{||}$, $J_{xz}$, and $J_H$, to investigate the pairing symmetry and superconducting gap structure on the FS in La$_3$Ni$_2$O$_7$ thin films.
Using exact diagonalization, the authors estimated $J^z_{\bot}\approx0.135$ eV.
The resulting superconducting gap projected on the FS is shown in Fig.~\ref{fig:5} (a), indicating a nodal $s\pm$ wave pairing symmetry.
The angular dependences of the gap on different pockets are shown in Fig.~\ref{fig:5} (b)–(d).
The gaps on the $\beta$ and $\gamma$ pockets are nodeless and have opposite signs, while a sign change occurs on the $\alpha$ pocket.
Whether there are nodes on the $\alpha$ pocket still needs more experimental results to prove.
On the $\beta$ pocket, the gap shows a parabolic shape centered around 45$^\circ$, and it slightly decreases as it nears the corner.
Orbital-resolved superconducting gaps are displayed in Fig.~\ref{fig:5} (e)–(h).
Qiu~\textit{et al.} systematically studied the phase patterns and pointed out that they can be interpreted as maximizing the overall gap magnitude on FS~\cite{Qiupairing2025}.
As a result, $\Delta_{||}^z$ tends to maximize the gap on the $\gamma$ pocket, whereas $\Delta_{||}^x$ enhances the gap on the $\alpha$ pocket, with only minor suppression on the $\beta$ pocket.
{\color{black} Moreover, as temperature rises, the energy gap of all pairing bonds decreases to zero at approximately 60 K in a mean-field manner.}
Recently, some researchers have paid attention to the prediction of the $T_c$ in superconductors~\cite{Qinintrinsic2025,Wanguniveral2025}.
They tried to explore intrinsic constraints on the $T_c$ in unconventional superconductors~\cite{Qinintrinsic2025} and give an empirical scaling relation connecting the maximum $T_c^*$ to the effective on-site Coulomb interaction $U$ in unconventional superconductors~\cite{Wanguniveral2025}. 
While the relation appears robust across some correlated superconductors, its microscopic origin remains an open question.


\section{\label{sec:6}Summary}
To date, some issues still need clarification. 
First, the role of the $\gamma$ pocket in enabling SC in bilayer nickelate thin films remains to be fully clarified.
Second, the relationship between the lattice ratio $c/a$ and the superconducting transition temperature $T_c$ in thin films differs from that in bulk La$_3$Ni$_2$O$_7$. 
These unresolved and seemingly contradictory results highlight the need for further experimental work to better understand SC in bilayer nickelate thin films.

\section{Acknowledgments}
We are grateful to Zhihui Luo and Cui-Qun Chen for useful discussions.
Work at Sun Yat-Sen University was supported by the National Natural Science Foundation of China (Grant Nos. 12494591, 92565303), the National Key Research and Development Program of China (Grant No. 2022YFA1402802), the Guangdong Provincial Key Laboratory of Magnetoelectric Physics and Devices (2022B1212010008), the Research Center for Magnetoelectric Physics of Guangdong Province (2024B0303390001), and the Guangdong Provincial Quantum Science Strategic Initiative (Grant No. GDZX2401010).

\bibliography{newref}

\end{document}